\documentclass{emulateapj}
\submitted{Submitted: \today}

\usepackage{amsmath}
\usepackage{natbib}
\setcitestyle{numbers,square}

\newcommand{\bfl}{\begin{flushleft}}
\newcommand{\efl}{\end{flushleft}}
\newcommand{\bea}{\begin{eqnarray}}
\newcommand{\eea}{\end{eqnarray}}
\newcommand{\be}{\begin{equation}}
\newcommand{\ee}{\end{equation}}
\newcommand{\bi}{\begin{itemize}}
\newcommand{\ei}{\end{itemize}}

\def\bec{\begin{center}}
\def\eec{\end{center}}
\def\beq{\begin{equation}}
\def\eeq{\end{equation}}

\newcommand{\ibar}{\overline{i}}
\newcommand{\jbar}{\overline{j}}
\newcommand{\kbar}{\overline{k}}
\newcommand{\lbar}{\overline{l}}

\newcommand{\nbar}{\overline{n}}
\newcommand{\Phibar}{\overline{\Phi}}
\newcommand{\Psibar}{\overline{\Psi}}

\newcommand{\p}{\partial}

\newcommand{\Xbar}{\overline{X}}
\newcommand{\ben}{\begin{enumerate}}
\newcommand{\een}{\end{enumerate}}

\begin{document}

\title{K\"ahler Moduli Inflation in Type IIB compactifications: A random tumble through the Calabi-Yau landscape}
\author{Richard Galvez}
\affil{Department of Physics and Astronomy, Vanderbilt University, Nashville, TN 37212, USA}
\email{Richard.A.Galvez@Vanderbilt.edu}

\begin{abstract}
In this paper we present an initial exploration of the Calabi-Yau landscape in the context of K\"ahler moduli inflation. We review how the slow-roll requirement on the scalar potential translates to a geometric constraint on the K\"ahler geometry of the vacuum. This constraint leads to a hard bound on the moduli space geometry and we consider the effects of this constraint on the string landscape that arises in type IIB string compactifications on an O3/O7 orientifold. Most notably we find that the inflationary constraint is independent of the moduli space dimension and only 6.57\% of geometries inspected support high-scale K\"ahler moduli inflation.
\end{abstract}


\section{Introduction}

An important question that has guided string phenomenology for over a decade is whether string theory can make predictive contact with the inflationary era. With the inflationary energy scale $H$ possibly being as high as ~$10^{16}$ GeV and beyond, inflation becomes a window into physics beyond the Standard Model of Particle Physics (BSM) and perhaps string theory. The hope for cosmological string phenomenology, or any BSM phenomenology in this direction, is that experimental measurements from the Cosmic Microwave Background (CMB) or other cosmological probes can restrict or guide phenomenological model building. 

In order to make contact with previous theoretical work in inflationary theory, typically two assumptions are made regarding the scale of inflation. First, the energy scale at which string effects become important is $M_{s} > H$ and thus inflationary dynamics are only sensitive to the massless states of the string. The governing theory in this case reduces to an effective supergravity theory in 10 dimensions. 

Secondly, the further assumption is made that 6 of the 10 dimensions are already compactified by the time the universe enters the inflationary era. That is, $H < M_{\rm kk}$ where $M_{\rm kk}$ represents the (possibly many) Kaluza-Klein scales introduced during the compactification of extra dimensions. While considering inflation from an extra dimensional perspective as is done in \citep{Brandenberger:1988aj} and \citep{Berera:2015yna} can be theoretically interesting and insightful, the majority of the literature has focused on the case where $H < M_{\rm kk} < M_s $ since the latter would require a rethinking of fundamental issues with the Hot Big Bang scenario such as the Horizon Problem, the Flatness Problem etc. \citep{Baumann:2014nda}.

After taking these two assumptions, one is left with an effective $\mathbb{N} = 1$ SO(3,1) supergravity theory whose dynamics still recalls its string-theoretic origins. The current state of the art in this direction can therefore be summarized quite simply by the following expression \citep{Baumann:2014nda}
\be
S_{10}[\mathbb{C}] \to S_4[\Phi[t]]
\ee 
where all of the geometric information of the compactification $\mathbb{C}$ is transferred to time-dependent fields $\Phi[t]$ in the 4D effective theory. In this perspective, the real and imaginary parts of $\Phi[t]_i$ account for all the moduli, fluxes, axions, etc, that arise due to the unique compactification $\mathbb{C}$. Unfortunately, as introduced in \citep{Douglas:2006es}, the overwhelmingly large number of possible compactifications $\mathbb{C}$ produce an unimaginably large number of vacuua\footnote{Typically the number $10^{500}$ is quoted \citep{Douglas:2006es}; for a popular review on this topic, see \citep{Bousso:2004fc}.}. 

This large number of fields $\Phi_i$ that appear in generic string compactifications provide for a rich phenomenology as the universe cools \citep{Acharya:2012tw} . They provide many candidate fields to drive inflation,  reheat the Standard Model, explain the dark matter relic abundance, source baryogenisis/leptogenisis and explain the matter/anti-matter asymmetry, etc. For an exhaustive review, see \citep{Baumann:2014nda}. 

However, this large number of $\Phi_i$ also presents a problem as there can be too many fields produced that cannot be explained physically, or worse, can spoil Big Bang Nucleosynthesis or over-close the universe. This problem, known as the cosmological moduli problem \citep{Banks:1993en}, has fortunately been shown to be resolved in various examples in generic string compactifications \citep{Randall:1994fr, Banks:1995dt, Acharya:2008bk}, giving hope that such an issue can be avoided in a realistic phenomenological model.

Slow roll inflation can also be sourced from the moduli and axion fields $\Phi_i$  \citep{Kachru:2003aw} . This phenomena is addressed in the scalar sector of an $N=1, D=4$ low-energy effective supergravity limit of the full UV complete theory. The scalar sector of the Lagrangian density therefore holds the relevant information for inflation and is given by
\be\label{eq:multifieldLagrangian}
\mathcal{L} = \frac{1}{2} R - g_{i \jbar} \p \Phi^i \p \overline{\Phi}^{\jbar} - V ( \Phi^i , \overline{\Phi}^{\jbar}  )
\ee
where $g_{i \jbar}$ is the metric of moduli space spanned by the scalars $\Phi$. In the SUGRA case $V ( \Phi^i , \overline{\Phi}^{\jbar}  )$ takes a special constrained form, while the form of $g_{i \jbar}$ contains the relevant geometric information of the string compactification. This is covered in detail in section \ref{sec:begining}.

Given a full string theoretic construction with stabilized moduli, the potential of the moduli fields $V$ should in principle be determined to enough accuracy to make contact with cosmological and astrophysical observations. Of course, this situation is notoriously difficult to achieve, given the complexity of such compactifications and the difficulty of fully incorporating quantum corrections. 

Nevertheless, the fact that the geometric information of $\mathbb{C}$ is encoded in the metric $g_{i \jbar}$ provides us a physical link between the string compactification geometry and the inflationary potential $V$. We may then ask the question if it is possible to restrict the landscape to those compactifications that support inflation.

Beyond the usual assumptions of the compactification and inflationary energy scales, we also make the assumption that the finite vacuum energy during inflation breaks supersymmetry. This assumption provides the physical link between the compactification geometry and the inflationary potential $V$: the {\it holomorphic sectional curvature} $\mathbb{H}[\Psi]$ along the Goldstino $\Psi$ direction \citep{Dutta:2012mw,Covi:2008cn}.

This purely geometric quantity that only depends on $g_{i \bar{j}}$ has to be larger than a specific value that depends on the Hubble scale and the Gravitino mass during inflation, with an absolute lower bound\footnote{This is in the case of low scale inflation where $m^2_{3/2} >> H^2$.}\footnote{A more general treatment of this bound is discussed in \cite{gonzalo}. Due to the lack of observational evidence for non-gaussianity, we restrict ourselves in this work to $c_s \sim 1$.} of 
\be \label{mastercondition}
\mathbb{H}[\Psi] > -\frac{2}{3},
\ee
where
\be \label{sectcurvcplx}
\mathbb{H}[\Psi]  \equiv  - \frac{R_{\Psi \Psibar \Psi \Psibar}}{g_{\Psi \Psibar} g_{\Psi \Psibar}}.
\ee

$\\$This condition is \textit{necessary} (but \textit{not} sufficient) for slow roll inflation and depends solely on the details of the K\"ahler geometry of the moduli space and not on the super potential. 

This allows one in principle to restrict geometries where the necessary condition is satisfied, before continuing in constructing a full model including a realistic super potential, or be concerned with an unrealistically high number of vacua. The aim of this paper and broader research program is to utilize the condition outlined in Eq. \ref{mastercondition} to restrict what should be considered to be part of the physical string landscape.

In section \ref{sec:distributionIntro} we outline several features of the distributions of holomorphic sectional curvatures that make this analysis particularly robust in the context of randomly sampled compactifications. We show that the distribution over many geometries and field values is independent of the moduli space dimension and the range of field values sampled.

In section \ref{sec:numerical} we use this approach to analyze physically relevant vacua in the context of K\"ahler inflation by sampling over sufficiently large compactification geometries and field values. We focus on the case of type IIB compactifications on O3/O7 orientifolds. In this case we find that 8.22\% of the compactifications studied support low-scale inflation, while only 6.57\% support high-scale inflation. Therefore, we note that the majority of compactifications inspected have a supergravity $\eta$ problem even at tree level and that the string landscape is barren of physical vacua supporting K\"ahler inflation.

In section \ref{sec:conclusions} we conclude and lay out a broader research program in this direction and future directions.  

\section{Holomorphic Sectional Curvature Limit from the CMB}\label{sec:begining}

For the purpose of exploring the inflationary landscape of type IIB compactifications in the geometric limit, we investigate the effective $D=4, ~\mathcal{N}=1$ supergravity (SUGRA) action. With chiral multiplet fields being denoted by $\Phi_I \equiv (\Phi_i,\Phi_{\ibar})$\footnote{We use the same notation for a superfield and its scalar component.}, the scalar potential solely depends on the K\"ahler potential $K$ and the super potential $W$, and their derivatives with respect to $\Phi_i$ and $\overline{\Phi}_{\overline{j}}$. We take $\{ \Psi_I \} \subset \{\Phi_I \}$ as the fields that acquire non-zero $F-$terms during inflation thereby breaking supersymmetry. 

The scalar fields $(\Phi_i,\Phibar_{\ibar})$ span a K\"ahler manifold $g_{i \jbar}$ which yields a scalar potential given by\footnote{We work in units where the Planck Mass $M^2=1$}
\be\label{pot:sugra}
V  =   e^K(g^{i\jbar} F_i F_{\jbar}  - 3|W^2|)\,\,,
\ee
where 
\bea
&F_i  =&  D_i W = \p_{i}W + \p_{i} K W \nonumber \\
&m_{3/2}^2 =& e^K |W|^2   .
\eea
While Eq. \ref{eq:multifieldLagrangian} describes a generic gravity theory coupled to scalars, Eq. \ref{pot:sugra} is the unique potential generated by a SUGRA theory. The K\"ahler geometry spanned by these theories provides extra complex structure to the space spanned by the scalars that non-supersymmetric theories would be completely void of. 

The metric, connection, and curvature tensor of a K\"ahler manifold $\mathcal{M}$ is given by
\bea
g_{i\jbar} &=& \partial_{i}\partial_{\jbar} K, \nonumber \\
\Gamma_{ij}^k &=& g^{\lbar k}\partial{i}g_{j\lbar}, \nonumber \\
R_{\kbar l \ibar j} = \p_{\ibar} \p_{j} \p_{l} \p_{\kbar} K \, &-& \, g^{\nbar m}(\p_{\ibar} \p_{\kbar} \p_{m} K)(\p_{j} \p_{l} \p_{\nbar} K)
\eea
where the key relation is the fact that the metric $g_{i\jbar}$ is given by two derivatives of the K\"ahler potential $K$.

Another geometric quantity of interest in the study of inflation is the the \textit{holomorphic sectional curvature} of a plane $(\Psi, \Psibar)$ defined in the tangent space at a given point in the manifold as \citep{Dutta:2012mw}
\be \label{sectcurvcplx}
\mathbb{H}[\Psi]  =  - \frac{R_{\Psi \Psibar \Psi \Psibar}}{g_{\Psi \Psibar} g_{\Psi \Psibar}}.
\ee
For a general theory of K\"ahler moduli inflation, this holomorphic sectional curvature $\mathbb{H}[\Psi]$ becomes intrinsically tied to the slow-roll condition \citep{Covi:2008cn} giving rise to a well defined bound on the underlying geometry to support inflation. 

To see this, we first consider the multifield potential slow-roll parameters given by \citep{Burgess:2004kv}
\bea
\epsilon \, &= \, \frac{\nabla^i V \nabla_i V}{V^2} \nonumber \\
\eta \, &= \, {\rm min \,\,\, eigenvalue} \, \{ N \} \,\,,
\eea
where 
\be \label{Hessian}
N \, = \, \frac{1}{V} \left( \begin{array}{cc}
\nabla^i \nabla_j V  &  \nabla^i \nabla_{\overline{j}} V  \\
\nabla^{\overline{i}}\nabla_j V & \nabla^{\overline{i}} \nabla_{\overline{j}}V   
\end{array} \right). 
\ee
The covariant derivative on the K\"ahler manifold is defined as
\be
\nabla_{i}f^k \equiv \p_{i}f^k + \Gamma_{ij}^k f^j  
\ee
for any vector $f^k$ on $\mathcal{M}$ where $\nabla_i$ is a covariant derivative with respect to the metric $g_{i \overline{j}}$.

For any given unit vector $u^I = (u^i,u^{\ibar})$, it can be shown that $\eta \, \leq \, u_I N^I_J u^J$ where $I = (i, \overline{i})$ and $J = (j, \overline{j})$. Choosing $u^I = (F^{\Psi}, F^{\Psibar})/(\sqrt{2}|F|)$ and evaluating the relevant covariant derivatives, one finds that 
\be\label{eq:CMBbound}
\eta \, \leq \, \eta_{\rm max} \, \equiv \,  \frac{2}{3\gamma} + \frac{1+\gamma}{\gamma}\mathbb{H}[\Psi] + \mathcal{O}(\sqrt{\epsilon}) \,\,,
\ee
where $\gamma \, = \, \frac{1}{3}\frac{V}{m^2_{3/2}} \, \sim \, \frac{H^2}{m^2_{3/2}}$ is the ratio of the inflationary energy scale to the SUSY breaking scale. 

In what follows, we drop all terms involving $\epsilon$ since $\sqrt{\epsilon} < \mathcal{O}(10^{-3})$. We next use the fact that the measured spectral index of curvature perturbations of the CMB is measured as $n_s \sim 0.968 \pm 0.006$ \citep{Ade:2015lrj} and relate this to $\eta$ via $n_s = 1+2\eta$ to arrive at $\eta_{\rm observed} \sim -0.01$. Therefore we find that $\eta_{\rm max} \geq -0.01$.

In solving inequality \ref{eq:CMBbound} for $\mathbb{H}[\Psi]$, we find the following necessary bound on the holomorphic sectional curvature along the SUSY breaking direction
\be \label{inflationfullcond}
\mathbb{H}[\Psi] \, \geq \, - \frac{2}{3}\frac{1}{1+\gamma} \,\,.
\ee
This relation is a direct link between the K\"ahler geometry of the underlining SUGRA theory and inflationary observables measured from the CMB. This is a {\it necessary} but not a {\it sufficient} condition for slow-roll inflation.

While this bound depends on $\gamma$, the dependance is not strong. We can see the allowable ranges of $\mathbb{H}$ by examining the two extremes of $\gamma$. With low scale inflation ($\gamma \ll 1$), the weaker bound of $\mathbb{H}[\Psi] \geq -2/3$ is realized while the case of high-scale inflation ($\gamma \gg 1$) yields the absolute bound of $\mathbb{H}[\Psi] \geq 0$.

\section{Distributions of Holomorphic Sectional Curvatures}\label{sec:distributionIntro}

Given a suitable potential that satisfies the slow roll conditions, it is of course possible to embed such a potential in supergravity - see for example \citep{Braun:2015pza, Kallosh:2010xz}. The embedding will satisfy the constraint on the sectional curvature, and typically locate the supersymmetry breaking field to a flat, decoupled sector.

In this paper we wish to examine the case of supergravity potentials one obtains from string compactifications. With the geometric bound from the CMB of the holomorphic sectional curvature well defined in inequality \ref{inflationfullcond}, we may now consider K\"ahler functions of a specific type. In this section we consider the case of heterotic string models with Calabi-Yau (CY) compactifications as an example (type IIB compactifications will be discussed in detail in section \ref{sec:taubasis}).

The moduli space manifold $X$ of heterotic CY geometries \citep{Braun:2015pza} have Hodge numbers $(h^{1,1}, h^{2,1})$ with a basis of divisors $S_{i}$, $i = 1, ..., h^{1,1}$ \citep{Conlon:2006gv} which sets the moduli space dimensionality as $dim(X) = h^{1,1}$. The complex K\"ahler moduli $t^{i}$ are defined from the Kahler form $J = t^{i} S_{i}$. 

In the absence of $\alpha^{\prime}$ and $g_s$ corrections, the K\"ahler potential for the moduli $t^i$ and axions $\bar{t}_i$ is given by
\be \label{Kahlerpot}
K \, = \, -3 \ln V ,
\ee
where $V$ is the moduli space spanned by the complex scalar moduli $t_i$ and $\bar{t}_i$. In the case of a CY 3-fold the volume $V$ is given by
\bea
V &=& \frac{1}{6} \int_X J \wedge J \wedge J  \\
    &=& \frac{1}{6} d^{i j k} (t_i - \bar{t_i}) (t_j - \bar{t_j}) (t_k - \bar{t_k}),
\eea
with the $d_{i j k}$ being the triple intersection numbers in the integral basis of the toric divisors\footnote{Because the metric of the moduli space is uniquely determined by the K\"ahler function, the set of $d_{ijk}$ determines the overall topology of the moduli space manifold.}. 

As a specific example of how the holomorphic sectional curvature can be useful in these constructions, we next consider the case where $h^{1,1} = 1$. This is the K\"ahler function studied in the Racetrack inflation models of \citep{BlancoPillado:2004ns} and takes the form
\be \label{simple1}
K = -3 \ln \left(t_1 - \bar{t}_1 \right ),
\ee
with $\mathbb{H}[t_1] = -2/3$  $\forall t_1$. Hence in this case, the bound is satisfied only for very low-scale inflation. 

The situation changes slightly when the K\"ahler potential takes the form
\be \label{Kahlercoset1}
K \, = \, -n \log \left( \frac{1}{6} (t_1 - \bar{t}_1) + X \Xbar \right),  \,\,\, \, n \in \mathbb{N} \,\,,
\ee
which defines maximally symmetric coset spaces \citep{GomezReino:2006wv} for which $\mathbb{H}[t_1] = \mathbb{H}[X] = -2/n$. This implies that inflationary scenarios based on these supergravity geometries face the $\eta$ problem when $n=3$ and supersymmetry breaking is dominated by the modulus $t_1$ or the field $X$ \citep{Kachru:2003sx}.

\subsection{Moduli dimension independence} \label{randomsugra}

As outlined in the introduction, we wish to approach the analysis of K\"ahler functions of the form of Eq. \ref{Kahlerpot} from a statistical perspective. We will consider K\"ahler geometries of this form with supersymmetry breaking during inflation dominated by moduli. In this section we wish to outline some general statistical results encountered when studying K\"ahler functions of the form of Eq. \ref{Kahlerpot}.

We first note that the K\"ahler function can be seen as a log of a random cubic polynomial, where the precise form of the polynomial is given when the intersection numbers $d_{ijk}$ and the moduli space dimensionality $dim(\mathcal{M})$ ($h^{1,1}$) are specified. 

For example, consider the form of the volume for the case of $h^{1,1} = 2$
\bea\label{eq:2dvol_generic}
V_2 &=& \alpha (t_i - \bar{t}_i )^3 +  \beta (t_1- \bar{t}_1)^2 (t_2 - \bar{t}_2)\\
   && + \gamma (t_1 - \bar{t}_1) (t_2 - \bar{t}_2)^2 + \delta (t_2 - \bar{t}_2)^3
\eea
where $\alpha, \beta, \gamma ~\rm{and} ~\delta$ are related to the appropriate geometric intersection numbers. We may consider a distribution of holomorphic sectional curvatures spanned by the parameter space $(\alpha, \beta, \gamma, \delta)$ and then study what subspace allows for slow roll inflation.

With random cubic functions of this general form we can first make some general statements about the holomorphic sectional curvatures derived for these cases. First consider the simplified case when
\bea\label{eq:genericVolume}
K &=& -n \ln V \\
V &\propto& \zeta(t_1-\bar{t}_1)^p + C.
\eea
This corresponds to a single term in the volume dominating with $C$ and $\zeta$ representing terms independent of $t_1$ and $\bar{t}_1$. By Eq.~\ref{sectcurvcplx}, one obtains 
\be\label{largeVolHSC}
\mathbb{H}[t_1-\bar{t}_1] = -\frac{2}{np}.
\ee
This result may be used to quickly verify whether or not inflation is possible within generic SUGRA theories by simply reading off the powers and pre log factor, for instance, in Eq. \ref{simple1}. 

This simple relation also leads to a surprisingly far-reaching consequence, namely
\\
\\
{\it the holomorphic sectional curvature distribution is independent of the dimensionality of the moduli space.} 
\\
\\
In order to see this, we first note that the result of Eq \ref{largeVolHSC} does not depend on the pre-factor $\zeta$ or the function $C$. Next we examine the volume in the case when $dim(\mathcal{M}) = 3$

\begin{eqnarray}\label{eq:3modvol}
V_3 =& ~~\alpha (t_1 - \bar{t}_1)^{3} + \beta (t_2 - \bar{t}_2)^{3} + \gamma (t_3 - \bar{\tau}_3)^{3}\nonumber \\
   +&~~\delta (t_1 - \bar{t}_1)(t_2 - \bar{t}_2)^2 + \epsilon (t_3 - \bar{t}_3)(t_2 - \bar{t}_2)^2 \nonumber\\
   +&~~\mu(t_2 - \bar{t}_2)(t_3 - \bar{t}_3)^2 + \nu(t_1 - \bar{t}_1)(t_3 - \bar{t}_3)^2 \nonumber\\
   +&~~\xi(t_2 - \bar{t}_2)(t_1 - \bar{t}_1)^2 + \theta(t_3 - \bar{t}_3)(t_1 - \bar{t}_1)^2 \nonumber\\
   +&\kappa (t_1 - \bar{t}_1)(t_2 - \bar{t}_2)(t_3 - \bar{t}_3) \nonumber\\~\nonumber\\
   \equiv&  \bar{\alpha} (t_i - \bar{t}_i )^3 +  \bar{\beta} (t_1- \bar{t}_1)^2 (t_2 - \bar{t}_2)\nonumber\\
   +&  \bar{\gamma} (t_1 - \bar{t}_1) (t_2 - \bar{t}_2)^2 + \bar{\delta} (t_2 - \bar{t}_2)^3 + C
\end{eqnarray}
with the pre-factors representing the appropriate geometric intersection numbers. Comparing this form of $V_3$ and $V_2$ of Eq \ref{eq:2dvol_generic}, we note that the volume in N-dimensional moduli space takes the same form. 

Therefore, the induction step follows 

\begin{eqnarray*}
V_N =& \frac{1}{6} d^{i j k} (t_i - \bar{t_i}) (t_j - \bar{t_j}) (t_k - \bar{t_k}) \\
   \equiv&  \bar{\alpha} (t_i - \bar{t}_i )^3 +  \bar{\beta} (t_1- \bar{t}_1)^2 (t_2 - \bar{t}_2)\\
   +&  \bar{\gamma} (t_1 - \bar{t}_1) (t_2 - \bar{t}_2)^2 + \delta (t_2 - \bar{t}_2)^3 + C,
\end{eqnarray*}
and hence the result of Eq.~\ref{largeVolHSC} holds for $dim(\mathcal{M}) = N ~\forall N \in \mathbb{Z}$. 

Following the result of Eq. \ref{largeVolHSC}, we thus expect the distributions of holomorphic sectional curvatures to be independent of moduli space dimension. We verify this result numerically in section \ref{sec:numerical}.

\begin{figure*}
     \begin{center}\hspace{-4pt}\includegraphics[scale=0.65]{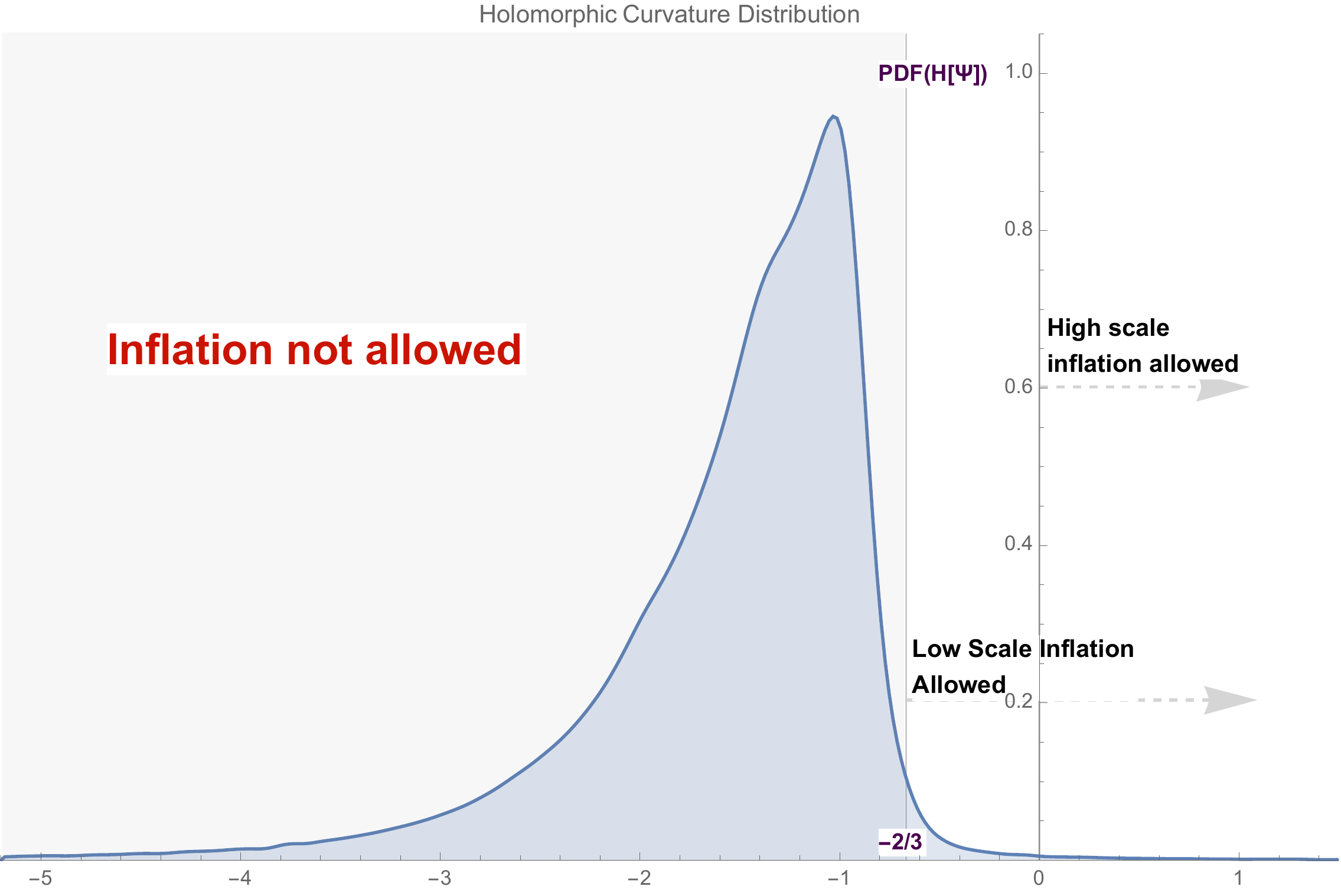}
     \caption{The distribution of $\mathbb{H}[\tau_1]$ for $h^{1,1}=2$ sampled over moduli values between $[1,12]$ and intersection numbers ranging between $[-8,8]$. We see the asymptotic values for $\mathbb{H}[\tau_1]$ dominating as expected.} 
     \label{fig:typeIIB2}\end{center}
 \end{figure*}
 
\begin{figure*}\vspace{-10pt}
     \begin{center}\includegraphics[scale=0.75]{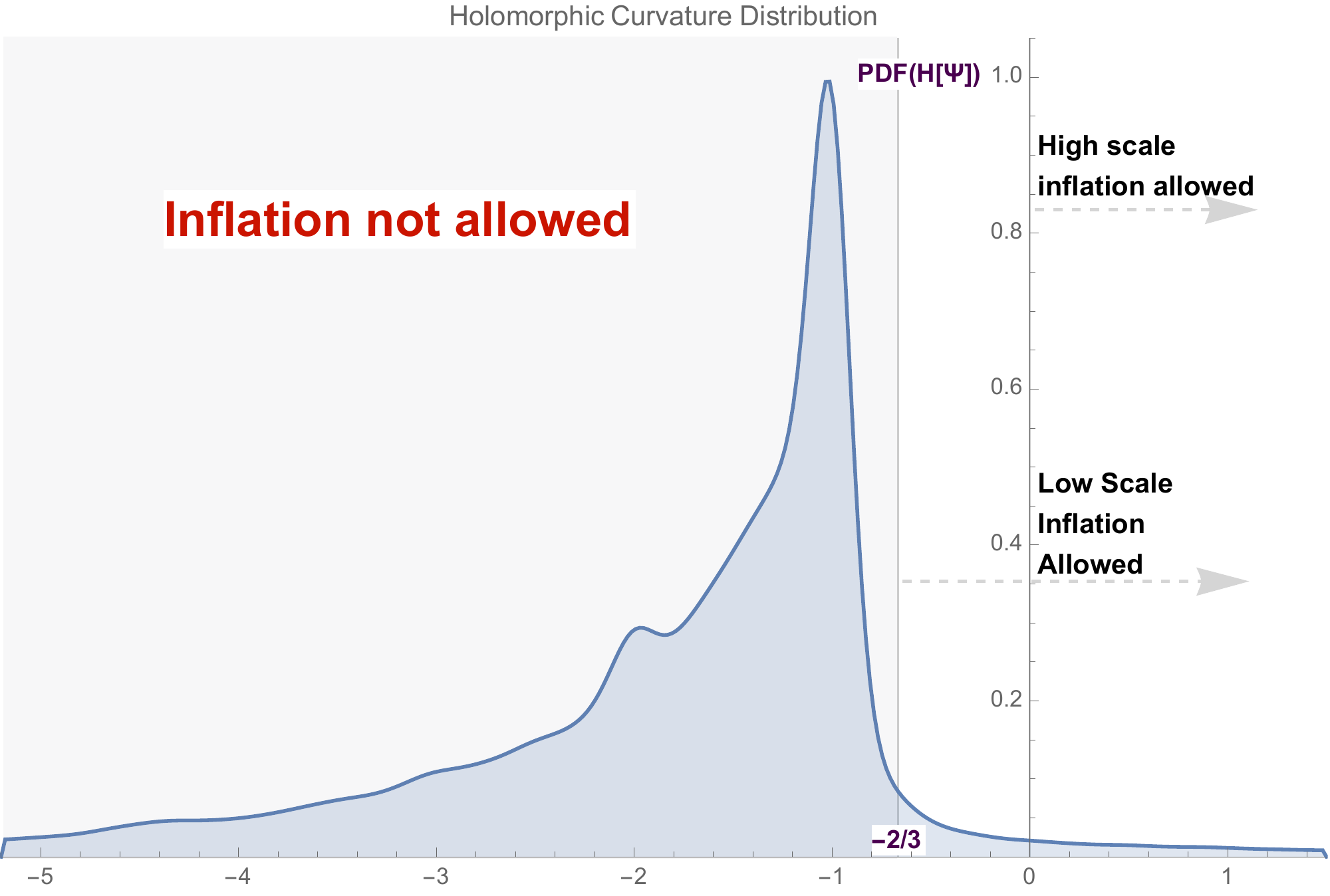}
    \caption{The distribution of $\mathbb{H}[\tau_1]$ for $h^{1,1}=3$ sampled over moduli values between $[1,5]$ and intersection numbers ranging between $[-3,3]$. We see the asymptotic values for $\mathbb{H}[\tau_1]$ dominating as in the case of $h^{1,1}=2$. The small peak at $-2$ is attributed to having a smaller sample size as compared to the case for $h^{1,1}=2$.} 
     \label{fig:typeIIB3}\end{center}
 \end{figure*}

\section{Type-IIB CY-3 compactifications on an O3/O7 orientifold}\label{sec:taubasis}

In our numerical study we focus on type-IIB CY-3 compactifications on O3/O7 orientifold involutions as it is a relatively well researched area \citep{Long:2014fba, Long:2016jvd}. In this case, the K\"ahler moduli space coordinates are the only unconstrained moduli as both the complex structure moduli and axiodilaton degrees of freedom can be fixed with appropriate choice of complex flux density \citep{Uranga:2004bw}. 

The holomorphic coordinates on the remaining moduli space are the complexified volumes of four cycles given by
\be
\tau_i \, = \, \int_{S_i} \frac{1}{2} J \wedge J \, - \, i C_4 
\ee
where $J$ is the complex K\"ahler form and $C_4$ the Ramond-Ramond four-form. The volume and hence the Kahler potential can be written implicitly in terms of the complex $\tau_i$ by using the relations
\be \label{ttotau}
\tau_i \, = \, \frac{\partial V}{\partial t_i} \, = \, \frac{1}{2} S_i J^2 \, = \, \frac{1}{2}d_{ijk} t^j t^k \,\,.
\ee

In type IIB, the supersymmetry breaking fields are the $\tau_i$ and hence the phenomenologically correct distribution to study is the distribution of $\mathbb{H}(\tau_i)$ for this case\footnote{In the heterotic case, the distributions turn out to be identical. This is due to dualities between type IIB and the heterotic string such that the inflationary observables are exactly the same in both cases.}. 

The Legendre transformation in Eq.~\ref{ttotau} is in general non-trivial, and obtaining an analytic expression for the K\"ahler potential in terms of $\tau_i$ given the intersection numbers $d_{ijk}$ is not straightforward. However, in considering random distributions one has the advantage that that the moduli space volume may be written as 
\bea
V &=& \frac{1}{6} d^{i j k} (t_i - \bar{t_i}) (t_j - \bar{t_j}) (t_k - \bar{t_k}) \\
   &\equiv& \frac{1}{6} c^{i j k} (\tau_i - \bar{\tau_i})^{1/2}(t_j - \bar{t_j})^{1/2} (t_k - \bar{t_k})^{1/2}
\eea
where the rational pre-factors $c^{ijk}$ are related to the intersection numbers $d^{ijk}$ via the Legendre transformation of Eq. \ref{ttotau}. 

This transformation allows one to utilize the methods outlined in section \ref{sec:distributionIntro} since the K\"ahler function may now simply be regarded as a homogeneous log polynomial of rank $3/2$ with random coefficients. Therefore, again in the case of $h^{1,1} = 2$, the volume may now be expressed in terms of $\tau_i$ as 
\begin{eqnarray}\label{eq:2dvol_generictao}
V_2 &=& \alpha (\tau_i - \bar{\tau}_i )^{3/2} +  \beta (\tau_1- \bar{\tau}_1) (\tau_2 - \bar{\tau}_2)^{1/2} \nonumber \\
   && + \gamma (\tau_1 - \bar{\tau}_1)^{1/2} (\tau_2 - \bar{\tau}_2) + \delta (\tau_2 - \bar{\tau}_2)^{3/2}
\end{eqnarray}
where the K\"ahler potential in the case of Type IIB is given by
\be\label{eq:tauK}
K = -2 \ln V.
\ee

This result is also independent of $h^{1,1}$ as was previously shown, however in the next section we verify this claim with $h^{1, 1} = 2, 3$. 
   
%
\section{Numerical results}\label{sec:numerical}

In this section we outline the numerical results obtained for the holomorphic sectional curvature distribution when uniformly sampling over geometric pre-factors $c^{i j k}$ and moduli $\tau_i$ for Eq. \ref{eq:tauK} with $h^{1,1} = 2, 3$. Due to the symmetries of the K\"ahler function, there will be no difference between the distributions of $\mathbb{H}[\tau_i]$ and $\mathbb{H}[\tau_j]$ therefore we focus on the distribution of $\mathbb{H}[\tau_1]$.
 
It is first useful to present the expectations from the asymptotic behavior of the holomorphic sectional curvature where we have substantial analytic handle on the result. To this end, we may use the result of Eq. \ref{largeVolHSC} to calculate the expected values of $\mathbb{H}[\tau_1]$ when various terms in the volume dominate. We therefore calculated the $\alpha, \beta, \gamma \to \infty$ first order corrections\footnote{We do not include the limit of $\beta \to \infty$ since $\mathbb{H}[\tau_1]$ is not defined for a term that does not include $\tau_1.$} separately of $\mathbb{H}[\tau_1]$ of Eq. \ref{eq:2dvol_generictao} for the case of $h^{1,1} = 2$, and arrive at

\be
\hspace{-20pt}\underset{{\alpha \to \infty}}{\mathbb{H}[\tau_1]} = -\frac{2}{3} + 
\frac{105\beta(\tau_2 - \bar{\tau}_2)^{3/2} -5\delta(\tau_1 - \bar{\tau}_1)\sqrt{\tau_2 - \bar\tau_2} }{24\alpha (\tau_1 - \bar{\tau}_1)^{3/2}}, \nonumber
\ee

\be
\hspace{-20pt}\underset{\delta \to \infty}{\mathbb{H}[\tau_1]} = -1 + \frac{9\alpha(\tau_1 - \bar\tau_1) - 45 \gamma(\tau_2 - \bar\tau_2)}{32\delta \sqrt{\tau_1 - \bar{\tau}_1}\sqrt{\tau_2 - \bar{\tau}_2}} , \nonumber
\ee

\be\label{eq:perturbativeH}
\hspace{-20pt}\underset{\gamma \to \infty}{\mathbb{H}[\tau_1]} = -2 + \frac{3\delta(\tau_1 - \bar\tau_1) - 15 \beta(\tau_2 - \bar\tau_2)}{8\gamma \sqrt{\tau_1 - \bar{\tau}_1}\sqrt{\tau_2 - \bar{\tau}_2}} , \nonumber
\ee
~\\~\\~
with similar results for Eq. \ref{eq:3modvol} when $h^{1,2} =3$.

\begin{figure*}
     \begin{center}\includegraphics[scale=0.65]{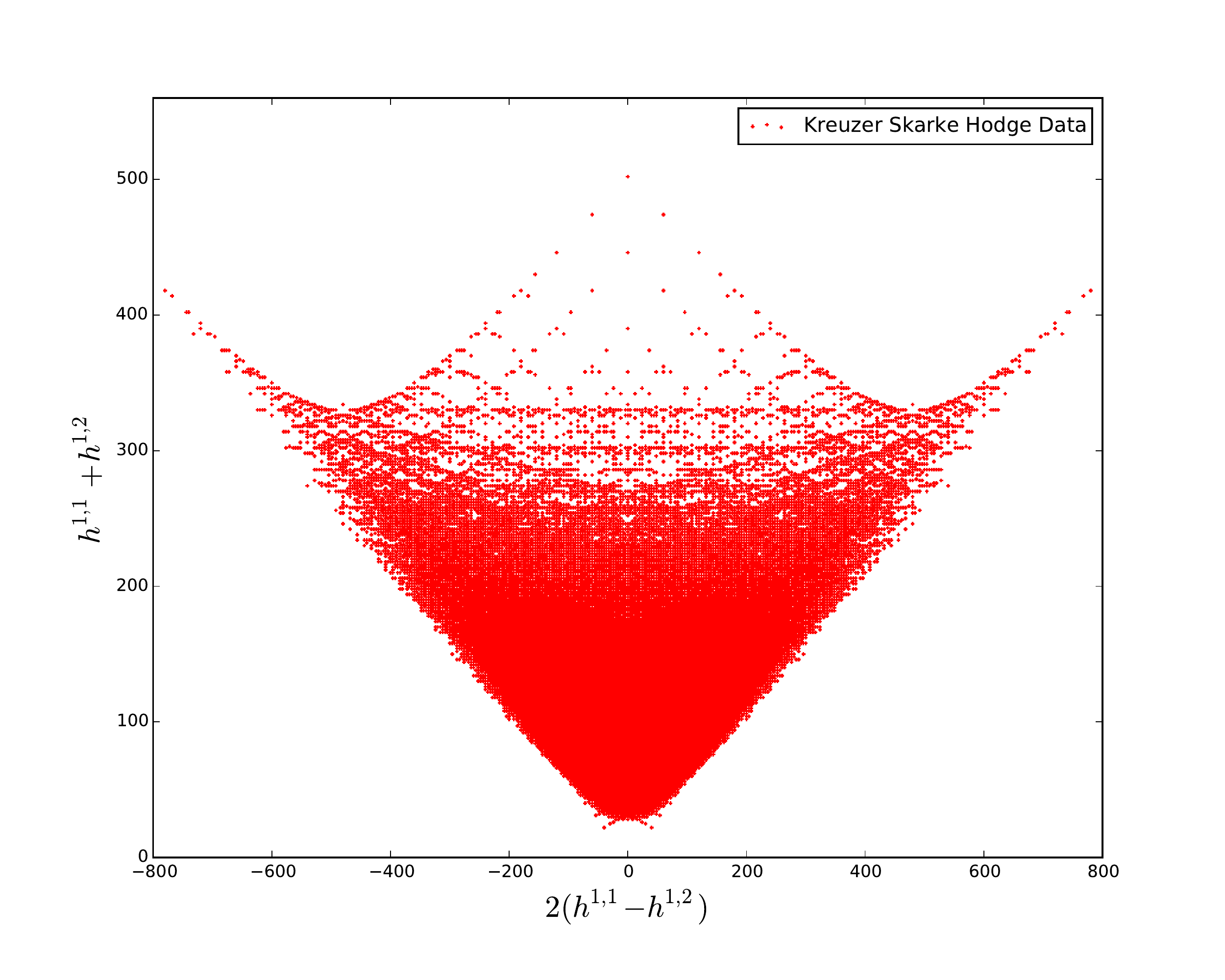}
     \caption{The Hodge plot of the Kreuzer Skarke database of 473,800,776 reflexive polyhedra that exist in four dimensions. The Hodge Numbers $h^{1,1}$ and $h^{1,2}$ are usually presented in this way to expose mirror symmetry. The x-axis is the Euler number of the compactification manifold.} 
     \label{fig:hodge}\end{center}
 \end{figure*}

From these results it follows that the distribution should be overwhelmingly shifted towards the negative while an interplay between the geometric coefficients and the four cycles interpolate between the first order values. These first order results could have readily been extracted off of Eq. \ref{eq:2dvol_generictao} using Eq. \ref{largeVolHSC}.
\be\label{eq:kahlergeneral}
\mathbb{H}[\tau_1] = -\frac{2}{np}  ~~~ {\rm with ~~} n = 2 {\rm ~~and ~~}2p = \{~1, 2, 3~\}
\ee
where $2p$ turns out to cycle through the complex dimension of the CY compactification $X$.

We then proceed to numerically calculate the distribution of $\mathbb{H}[\tau_1]$ by sampling over 
\be
\tau_i \in [1,12] ~,~ {\rm ~and~~}C^{i j k} \in [-8,8]
\ee
for $h^{1,1} =2$ and
\be
\tau_i \in [1,5] ~,~ {\rm ~and~~}C^{i j k} \in [-3,3] 
\ee
for $h^{1,1} =3$ and build a probability density function (pdf) for both cases. The volume sampled for $h^{1,1} = 3$ is that of Eq. \ref{eq:3modvol}. The pdf obtained for $h^{1,1} = 2$ is presented in Fig. \ref{fig:typeIIB2} while the pdf for $h^{1,1} = 3$ is presented in Fig. \ref{fig:typeIIB3}. This numerically verifies our claim of $h^{1,1}$ independence on the holomorphic sectional curvature distribution.

In analyzing these distributions we see that there is a strong preference for $\mathbb{H}[\tau_1] =-1$ in both cases. This can be understood by analyzing the perturbative results of Eq. \ref{eq:perturbativeH}. The first order result of $\mathbb{H}[\tau_1] $ as $\delta \to \infty$ shows that the first order correction has the largest pre-factor suppression of $\frac{9}{32}$ and $\frac{45}{32}$ and hence values of $\mathbb{H}[\tau_1] \sim-1$ are prominent in the distribution. 

One result that stands out from this numerical study is that a random exploration of geometric numbers $C^{ijk}$ leads to an overwhelmingly large set of vacua that cannot support K\"ahler inflation. In particular, we find that 8.22\% of the compactifications support low-scale inflation, while only 6.57\% of the compactifications allow for high-scale inflation.

\section{conclusion and future work}\label{sec:conclusions}
Our main results can be outlined as follows:
\\
$(i)$ the distribution of holomorphic sectional curvatures, and hence the likelihood of K\"ahler moduli inflation occurring, is independent of the dimensionality of the moduli space. 

This is due to the structure of the moduli space volume being a generic polynomial of degree $p$, regardless of the number of variables. This makes it clear that this approach is orthogonal to previously studied methods \citep{Marsh:2011aa} which rely crucially on the presence of a large number of moduli fields.

$(ii)$ The distribution is independent of the range $\tau_i$, as long as the range is democratic for all moduli fields. This can be attributed to the fact that the K\"ahler potential is a log of a homogeneous polynomial, and $\mathbb{H}$ only depends on derivatives of the K\"ahler potential.

$(iii)$ We find that 8.22\% of the compactifications studied support low-scale inflation, while only 6.57\% of the compactifications allow for high-scale inflation. The vast majority of the landscape therefore does not support K\"ahler moduli inflation.

\subsection{Future work}

The independence on moduli space dimension makes the study of a distribution over geometries exceptionally general. This is of course considering the case of a uniform distribution of $C^{ijk}$. However, given a set of $C^{ijk}$ it is not guaranteed that one has a Calabi-Yau manifold since $C^{ijk}$ have to be chosen carefully for such to be the case, see for instance \citep{Denef:2004dm}. Hence, sampling over the physical Calabi-Yau landscape, as opposed to a random sampling as was done in this work, is a more complex matter.

To this end, we next focus our attention to studying the largest database of Calabi-Yau compactifications currently known produced by Kreuzer and Skarke in their famous construction of all 473,800,776 reflexive polyhedra that exist in four dimensions\footnote{A clear review is presented in \citep{Altman:2014bfa} where the extraction of physically relevant topological data is emphasized, such as the intersection numbers $d^{ijk}$ of Calabi-Yau compactifications.} \footnote{For an alternative exploration of Calabi-Yau constructions in 3 dimensions, see \cite{Gao:2013pra}.} \citep{Kreuzer:2000xy}. The Hodge plot of all of these Calabi-Yau compactifications is presented in Fig. \ref{fig:hodge} and is the best known evidence for the existence of mirror symmetry. See \cite{He:2015fif} for an initial exploratory analysis of the KZ dataset and \cite{Sinha:2015zva, Easther:2013nga} for a broad review on phenomenological implications.

We propose to study this database with the holomorphic sectional curvature distribution methods outlined in this paper. We wish to consider if a uniform distribution of geometric numbers $C^{ijk}$ gives rise to a different distribution than that of the KZ database, and if similar percentages of the landscape are excluded. We leave this for future work. 

 \section*{Acknowledgements}
RG is grateful to Kuver Sinha, Thomas Grimm, Scott Watson, Liam McAllister, Cody Long, Paul McGuirk, John Kehayias, Robert Scherrer, Tom Kephart and Keivan Stassun for useful and insightful discussion. This work was supported by NSF PAARE grant AST-1358862.
\appendix
\section{Extensions to higher dimensional Calabi Yau compactifications}
The compactification dimension $dim(\mathcal{M})$ is typically fixed to 3 complex dimensions to achieve a critical string theory and have physical contact with our universe. As a mathematically curious result however, we may extend this analysis to $m$ complex dimensional Calabi-Yau compactifications. For instance, elliptically fibered CY-4 folds are often studied in the construction of F-theory compactifications \citep{Vafa:1996xn} whereby one can embed a CY complex structure on the moduli space of 10 dimensional type II-B string theories after asymptotically shrinking away one of the complex dimensions\footnote{Often referred to as the F-Theory limit.}.  

For general CY-$m$ manifolds, the moduli space volume is defined as
\be
V_m = \frac{1}{m!} d^{i_1 i_2 \ldots i_m} (t_{i_1} - \bar{t_{i_1}})(t_{i_2} - \bar{t_{i_2}}) \ldots (t_{i_m} - \bar{t_{i_m}})
\ee
and the K\"ahler function defined as
\be
K = -n \log{V_m}. 
\ee
The result of Eq. \ref{eq:kahlergeneral} is immediately extended to the logs of random polynomials of degree k with $0^{th}$ order values of
\be \label{peakstructureN}
\mathbb{H}[\tau_i] = -\frac{2}{nk}  ~~~ {\rm with ~~} 2k = \{1, 2, \ldots ,m \}
\ee
allowing one to analyze higher dimensional CY geometries, if desired.

\newpage
\bibliography{references}

\end{document}